\RequirePackage{fix-cm}
\documentclass[smallextended]{svjour3}       
\smartqed  
\usepackage{appendix}
\usepackage{amsmath}
\usepackage{graphicx}
\usepackage{lineno}
\usepackage{array}
\usepackage{longtable}
\usepackage{natbib}
\usepackage{pgfplots}
\usepackage{caption}
\usepackage{subcaption}
\usepackage{algorithm}
\usepackage{algpseudocode}
\usepackage{tabularx}
\usepackage[T1]{fontenc}
\usepackage[utf8]{inputenc}
\usepackage{newtxtext,newtxmath}
\usepackage{latexsym}
\usepackage{multirow}
\usepackage{threeparttable}
\usepackage{float}
\usepackage{flushend}
\usepackage{adjustbox}
\usepackage[bookmarks=false]{hyperref}
\hypersetup{
	colorlinks=true,
	linkcolor=blue,
	filecolor=blue,      
	urlcolor=blue,
	citecolor = blue
}
\usepackage{float}
\usepackage{hyperref}
\newcommand*\patchAmsMathEnvironmentForLineno[1]{%
\expandafter\let\csname old#1\expandafter\endcsname\csname #1\endcsname
\expandafter\let\csname oldend#1\expandafter\endcsname\csname end#1\endcsname
\renewenvironment{#1}%
{\linenomath\csname old#1\endcsname}%
{\csname oldend#1\endcsname\endlinenomath}}%
\newcommand*\patchBothAmsMathEnvironmentsForLineno[1]{%
\patchAmsMathEnvironmentForLineno{#1}%
\patchAmsMathEnvironmentForLineno{#1*}}%
\AtBeginDocument{%
\patchBothAmsMathEnvironmentsForLineno{equation}%
\patchBothAmsMathEnvironmentsForLineno{align}%
\patchBothAmsMathEnvironmentsForLineno{flalign}%
\patchBothAmsMathEnvironmentsForLineno{alignat}%
\patchBothAmsMathEnvironmentsForLineno{gather}%
\patchBothAmsMathEnvironmentsForLineno{multline}%
}

\begin{document}

\title{RikoNet: A Novel Anime Recommendation Engine}


\author{Badal Soni \and Debangan Thakuria \and Nilutpal Nath \and Navarun Das \and Bhaskarananda Boro       
        }


\institute{Badal Soni  \at
              National Institute of Technology Silchar, India \\
         \email{badal@cse.nits.ac.in}           
           \and
           Debangan Thakuria \at
             National Institute of Technology Silchar, India 
             \\
     \email{debanganthakuria44@gmail.com}  
              \and
           Nilutpal Nath \at
             National Institute of Technology Silchar, India 
             \\
     \email{nilutpalnath555@gmail.com} 
      \and
           Navarun Das \at
             National Institute of Technology Silchar, India 
             \\
     \email{dasnavarun06@gmail.com}
     \and
          Bhaskar Boro \at
             National Institute of Technology Silchar, India 
             \\
     \email{boro123bhaskar@gmail.com}
            }

\date{Received: DD Month YEAR / Accepted: DD Month YEAR}

\maketitle

\begin{abstract}
Anime is quite well-received today, especially among the younger generations. With many genres of available shows, more and more people are increasingly getting attracted to this niche section of the entertainment industry. As anime has recently garnered mainstream attention, we have insufficient information regarding users' penchant and watching habits. Therefore, it is an uphill task to build a recommendation engine for this relatively obscure entertainment medium. In this attempt, we have built a novel hybrid recommendation system that could act both as a recommendation system and as a means of exploring new anime genres and titles. We have analyzed the general trends in this field and the users' watching habits for coming up with our efficacious solution. Our solution employs deep autoencoders for the tasks of predicting ratings and generating embeddings. Following this, we formed clusters using the embeddings of the anime titles. These clusters form the search space for anime with similarities and are used to find anime similar to the ones liked and disliked by the user. This method, combined with the predicted ratings, forms the novel hybrid filter. In this article, we have demonstrated this idea and compared the performance of our implemented model with the existing state-of-the-art techniques.
\end{abstract}
 

\section{Introduction}

\label{intro}
\emph{Anime} (a term derived from the English word \emph{animation}) is a form of hand-drawn computer animation which originated in Japan and has now developed a cult following around the world. In recent years, the \emph{Anime} industry has been growing at an enormous pace making billions of dollars in profit every year. Its market has gained attention from major streaming platforms like \href{https://www.netflix.com/}{Netflix} and \href{https://www.primevideo.com/}{Amazon Prime}. 

In the pre-internet era, \emph{Anime} enthusiasts discovered new titles through word of mouth. Hence personalized recommendations were not required. Moreover, the number of titles released were quite less to facilitate a data-based approach for personalized recommendations. However, in recent years, with the boom of streaming services and the amount of newly released anime titles, people can watch \emph{Anime} as much as they like. This calls for a personalized recommendation system for this new generation of \emph{Anime} watchers.

A recommendation system seeks to recognize a user's preferences and predict the items that are most likely to be rated highly by the same. Different filtering approaches are used by recommendation engines such as content-based, collaborative filtering, multi-criteria, risk-aware, and hybrid filtering. The later being an ensemble of multiple other filtering techniques. 

The content-based filtering uses the features of the items and a user's profile created from the data of previous activities to generate recommendations. This technique is most appropriate when known information about the items (name, area, portrayal). In content-based recommender system, recommendations are treated as a user-specific classification problem where the classifier is trained based on the likes and dislikes of the user and item features. Items are described using keywords, and a client profile is created to store the kind of items the client likes. These algorithms use this user profile to recommend items similar to those the client is currently examining or enjoyed in the past. 

Collaborative filtering algorithms extract data from the behavior of users for recommending items. The exploitation of other users' behavior, such as transaction history, ratings, selection, and purchase information, is also a part of the process. New users are recommended items based on the preferences and behavior of other users. Collaborative filtering assumes that users who have converging opinions on a set of items in the past will probably like the same items in the future. 

A hybrid filter is the most commonly used approach for recommendation systems nowadays. It is formed by combining collaborative filtering, content-based filtering, and other approaches. A Hybrid approach combining collaborative and content-based filtering can be implemented in several ways: combining the predictions from content-based and collaborative-based filtering, which were separately made previously, by adding capabilities of one approach to another or by unifying the approaches into one model. Many studies comparing the performance of the hybrid method with the pure collaborative and content-based methods have shown the superiority of the hybrid method, which can provide more accurate recommendations. Prevalent problems in recommendation systems like the cold start problem, the sparsity problem, and the knowledge engineering bottleneck can be vanquished using the hybrid approach.

This article presents our \emph{RikoNet: A Novel Anime Recommendation Engine} prototype that combines the content-based and collaboration filtering approaches into one robust and effective recommendation system. We have extracted the content information from the \emph{genre} information and the users' interactions with the various anime titles. This information is used to cluster the anime titles into coherent clusters of similar items. Conversely, we have used an autoencoder for the collaborative filtering part, the details described in the sections below. This setup ensures that users' preferences are exploited along with an exploration of other genres of \emph{Anime}, thus broadening the horizons of the user. From the perspective of evaluation, the individual components of our RikoNet perform better than their corresponding counterparts. A study with 20 people has shown that the perceived quality of the recommendation lists generated by the overall system improves as the users interact more with the system. 

\subsubsection{Contributions}
The contributions of this article are apparent in the various aspects of our hybrid anime recommender system:
\begin{itemize}
    \item \emph{\textbf{Hybrid Recommender Framework:}} We have used a two-pronged approach for recommendation - to recommend animes similar to the users' current preferences, as well as animes which the user may like based on other similar users. This novel hybrid filter can help the user to follow his interests and explore new genres.
    \item \emph{\textbf{Deep autoencoders:}} The deep autoencoders that we have used for predicting user ratings perform better than the current state-of-the-art models and algorithms. This deep network can detect complex relationships among various genres of anime and different users.
    \item \emph{\textbf{User Interactive System:}} The users can continuously interact with our system, rating animes as they watch them. This enables RikoNet to adapt to the preferences of the user and generate personalised recommendations. 
\end{itemize}

\subsubsection{Article Organisation}	
The rest of the article is organised as follows. The Section \ref{Related Works} contains the related works which we have referenced during our research. Following which, we present our main methodology in Section \ref{Methodology}. Under this section, we discuss the main architecture and design of our RikoNet and the user-system interaction in Sections \ref{Arch} and \ref{System Interaction} respectively.

In Section \ref{Results}, we mention our experimental setup - the dataset and the performance parameters we used to evaluate our system. We also discuss how we pre-processed the dataset in details. Then we move on to the discussion and analysis of the results in Section \ref{Discussions}. Here we present a comparative evaluation of RikoNet first, followed by the results of “human evaluation". The experimental setup used for the later is also discussed here. Finally, we conclude our article in Section \ref{Conclusion}.
	
\section{Related Works} \label{Related Works}
In recent years there have been several works on recommendation systems. For instance, \cite{kuchaiev2017training} proposed a model named DeepRec for the rating prediction task in a recommender system, based on a deep autoencoder of 6 levels. They suggested that going deeper helps generalization and produces better recommendations. A new training algorithm based on iterative output re-feeding was also proposed to overcome the natural sparseness of collaborative filtering. However, unlike our model, the potential of their model is limited by the absence of content-based filtering.

Similar to our model, \cite{zhang2015hybrid} too came up with a hybrid recommender system based on user-recommender interaction. The sequence of the user’s choice affects the subsequent recommendation in the process of user-recommender interaction. Thus, the interaction forms an active learning scenario. The hybrid algorithm comprised of both \emph{random} and \emph{k}-nearest neighbors (\emph{k}NN) algorithms.

To handle the cold start problem effectively, \cite{ramashini2018personalized} designed their personalized recommender system for recommending leisure time activities to users, by using both collaborative filtering and content-based filtering on the social media data of the users. But on the other hand, it presented possible privacy and security risks. Our recommender system, Rikonet, does not present any such issue.

\cite{hornung2013evaluating} presented TRecS (Track Recommender System), a music recommender system with a motivation of introducing users to new music, apart from those from the user’s past listening history, that they may like; similar to what we are trying to achieve through Rikonet. The hybrid approach used comprises one collaborative filtering approach and two content-based ones. The three sub-recommenders contribute to the eventual prediction and the recommendation generation process.

\cite{beherahybrid} proposed a new hybrid recommender system to achieve more accurate predictions. In the proposed model, first, K-means algorithm was used for clustering the sample. Then, an SVM classifier was used on the dataset. Finally, the PSO algorithm was used to fine-tune the profile matching process to provide more accurate results.

\cite{barkan2016item2vec} proposed a neural embedding algorithm for item-based collaborative filtering- Item2Vec, which deals with the overlooked task of learning item similarities by embedding items in a low dimensional space regardless of the users. Motivated by the success of Skip-gram with Negative Sampling (SGNS), known also as Word2Vec \citep{mikolov2013distributed} in other domains, they designed a modified version of SGNS for the model. The quantitative and qualitative evaluations demonstrated the effectiveness of Item2Vec when compared to an SVD-based item similarity model.

Like Rikonet, \cite{kiran2020dnnrec} used deep learning to provide better recommendations from their recommender system- DNNRec. The deep neural network designed was of three layers, where the system used embeddings for representing users and items to learn non-linear latent factors. It could also alleviate the cold start problem by integrating side information about users and items into a very deep neural network.

\cite{kumar2015movie} proposed a movie recommendation system named MOVREC. It is based on a collaborative filtering approach that makes use of the information provided by users, analyses them, and then recommends the movies that are best suited to the user at that time. The system requires the user to select/input values from the set of movie attributes, after which a movie list is recommended based on the cumulative weight of different attributes and using the K-means clustering algorithm.

\cite{handeinternational} also took the hybrid approach between collaborative filtering and content-based filtering for designing their movie recommender system- Moviemender, to overcome the drawbacks of each algorithm. Content-based predictions were used to train each user-rating vector in the user-rating matrix and converted it into a pseudo rating matrix that combines actual rating with the predicted ratings. Collaborative filtering was then applied to this full pseudo-user-rating matrix to make recommendations for an active user. Clustering was also used for better recommendations, thus reducing MAE and increasing precision and accuracy.

\cite{vie2017using} used a CNN (Convolutional Neural Network) to extract tag information from the manga and anime posters to provide extra information for their proposed collaborative model - BALSE (Blended Alternate Least Squares with Explanation) to recommend mangas. The model relied on the rating matrix and the posters whenever the rating information was insufficient. This unique idea helped them tackle the item cold-start problem and improve the recommendation performance for the little-known items.

\cite{ota2017anireco} proposed AniReco: Japanese Anime Recommendation System that adopted a content-based filtering approach. The system used a network diagram based on the calculation of user’s preferences from the history of watching works and evaluated them using information regarding constituent elements of animation works. 

\cite{geetha2018hybrid} took a mixed approach as they combined collaborative filtering and content-based filtering approaches in their recommendation system, thereby improving the performance and accuracy of the system. The system allowed a user to select choices from a given set of attributes and then recommend him a movie list based on the cumulative weight of those attributes and K-means algorithm application.

\cite{cintia2020design} compared performances of several clustering algorithms, namely, K-Means algorithm, birch algorithm, mini-batch K-Means algorithm, mean-shift algorithm, affinity propagation algorithm, agglomerative clustering algorithm, and spectral clustering algorithm to obtain the nearest neighbor of the same cluster for the user. Then the recommendation system would finally recommend the list-N of movie list based on user similarity.

To overcome the limitation of overspecialization in content-based recommender systems, \cite{kim2011recommender} proposed the application of interactive evolutionary computation with it. It enabled their recommender system to adapt and respond to immediate changes in preferences effectively. Recommendations improved through the K-means algorithm.

Using a dataset obtained from Kaggle \cite{girsang2020collaborative}, proposed a recommendation system using collaborative filtering to recommend anime. Each user needed to add anime to their completed list and give it a rating to produce a data set for that user. Rather than clustering algorithms, the system used the SimRank tool for calculating the similarity of users in the collaborative filtering model.

\cite{al2016automated} presented a new collaborative recommendation system that employed an association rules algorithm to recommend university elective courses. The system generates course association rules that satisfy the support and confidence constraints entered. Then based on similar students, courses are recommended.

\cite{azfar2020evaluation} used three hybrid combinations- Cosine Similarity with k-Nearest Neighbors (KNN), Term Frequency-Inverse Document Frequency (TF-IDF) with Singular Value Decomposition (SVD) matrix factorization, and k-Means clustering with Jaccard similarity to provide recommendations. The system recommends the top 5 movies similar to the picked movie from the list of movies displayed based on the hybrid algorithm chosen.

\cite{napoles2020recommender} built a recommender system based on Long-term Cognitive Networks (LTCNs) that allows experts to inject knowledge into the network along with a three-step learning procedure. The architecture expands the LTCN model by adding Gaussian kernel neurons that compute estimates for the missing ratings of the user-item matrix and make predictions. However, since their approach is context-free and does not involve human experts, the system fails for new trends and may need human intervention.

Like RikoNet, \cite{virk2015analysis} proposed a hybrid recommender system based on content and collaborative filtering. The system used contextual information to cope with sparsity and scalability issues inherent in the collaborative filtering approach for better movie recommendations. Every new user needs to register with personal data and the ratings of a set of movies to solve the cold-start problem.

\section{Methodology} \label{Methodology}
\subsection{RikoNet Architecture and Design} \label{Arch}
\begin{figure}
    \centering
    \includegraphics[width=\textwidth]{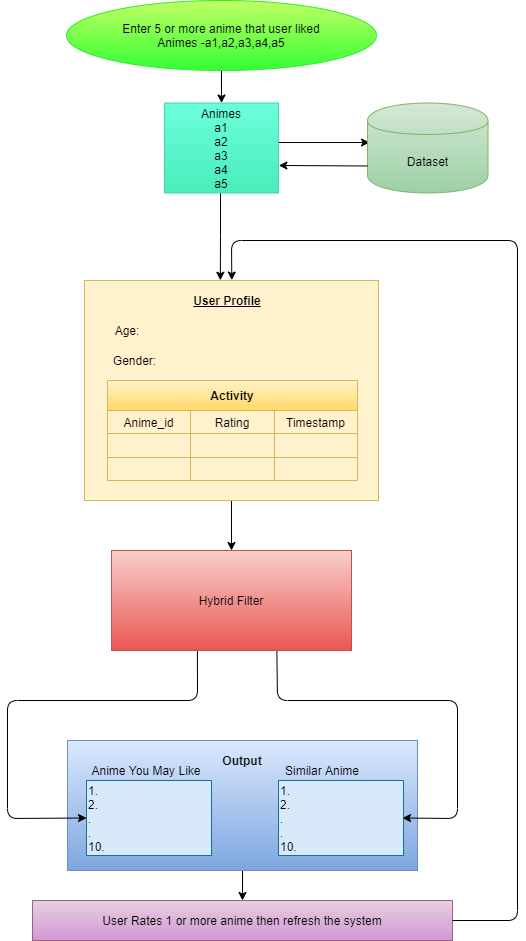}
    \caption{Proposed System Flow Diagram}
    \label{fig:mainpath}
\end{figure}

RikoNet relies mainly on an autoencoder, which takes in input the ratings provided by a new user to predict the ratings of the rest of the anime titles in our dataset. We have discussed the architecture of this autoencoder in Section \ref{Primary Autoencoder}. The overall architecture and design of the recommendation engine is shown in Fig. \ref{fig:mainpath}. Our crawled dataset from MyAnimeList.net is first pre-processed(e.g. data cleaning and disambiguation) and reduced to 6668 anime titles while retaining all the key information. Following that we used autoencoders to learn embeddings of all the anime titles present in our dataset, which were then used to cluster the same. The results of this process was pre-computed and stored in the knowledge base. The logically opposite clusters of the anime titles are estimated as well and added to the knowledge base. The algorithms used in this process are discussed in details in Section \ref{Logically Opposite Cluster}. At run-time, when a user requests a new recommendation list, the user's context, i.e., the anime titles rated so far is fed into the primary autoencoder, which computes the predicted ratings for the unrated titles. These ratings are further fed to a hybrid filter, which generates 2 lists, namely - \emph{Similar Anime} and \emph{Anime You May Like}, the former showing anime titles similar to the ones the user rated highly and the later showing titles which the user may like based on his overall ratings.

The prototype of RikoNet is available online with an introductory tutorial on how to obtain anime recommendations, in our github repository  \url{https://github.com/NilutpalNath/RikoNet}.

\subsubsection{Primary autoencoder}\label{Primary Autoencoder}

The autoencoder implemented in this paper is inspired from the one implemented in the paper \emph{Training deep autoencoders for collaborative filtering}, \cite{kuchaiev2017training}.

Autoencoder is a network which first encodes the n-dimensional inputs into a d-dimensional representation and then decodes it back to n-dimensional output. The transformations applied are $encode (x): R^n \rightarrow R^d$ and 
$decode (z): R^d \rightarrow R^n$. Here $R^x$ represents a vector of x dimensions.

The autoencoder is implemented as a feed forward neural network comprising of fully
connected layers. It is aimed at obtaining a d-dimensional representation of the input such that the error between \emph{x (original input)} and 
\emph{f(x) (produced by the network)} is minimum. 
The fully connected layers compute 
$z = f(W\times x + b)$, where f is a non-linear activation function. We
have used \emph{Scaled Exponential Linear Unit (SELU)} as the activation function for the hidden layers and ReLU for the final layer activation. Below are the mathematical formulation of the functions.
\begin{equation}
    SELU:\   g(x) = \lambda \left\{\begin{matrix}
    \alpha e^x - \alpha,& x<0\\ 
    x,& x\geq 0 
    \end{matrix}\right.
\end{equation}

\begin{equation}
    ReLU:\  g(x)=\left\{\begin{matrix}
    0& x<0\\ 
    x& x\geq 0
    \end{matrix}\right.
\end{equation}


The autoencoder is fed ratings for all the anime titles given by a user, keeping the ratings for unrated anime as 0. Using the autoencoder in this way, the predicted ratings for the corresponding user on all the unrated anime are obtained.

\subsubsection{Generate Embedding}

Top users are selected from the dataset who have rated a large number of anime. A threshold, \emph{t}, is selected for filtering out the top raters whose total number of ratings is greater than \emph{t}. A vector is created for each anime comprising of the ratings from the top users for the corresponding anime. All the anime vectors are of constant dimensions \emph{m}, which equals the number of top users. These vectors are fed into another autoencoder of the same architecture for training it. This autoencoder produces a compressed \emph{d-dimensional} encoding of each anime.
\begin{equation}
    encode(x): R^m \rightarrow R^d
\end{equation}

All the anime are one-hot encoded based on genres creating a \emph{g-dimensional} anime vector. It is appended to the already generated d-dimensional encoding of corresponding anime which adds knowledge of the contents of the anime to the vector representations. The vectors are thus transformed as, $R^d \rightarrow R^{d'}$, where $d' = d + g$.

The new vectors are used to train yet another autoencoder. Unlike the earlier ones, no final layer activation function is used in this autoencoder. Moreover, MSE without the \emph{masking feature} is used as the evaluation metric. This is done to train the autoencoder to reproduce back the exact input vector.The autoencoder yields the \emph{final hybrid encoding} of anime which are again set to be of \emph{d-dimensions}. The transformation which occur is $encode(x): R^{d'} \rightarrow R^d$.

\subsubsection{Clustering of Anime Titles}
The embedding of the anime titles are used to cluster them into appropriate categories. We use the \emph{Spectral Clustering} algorithm for our purpose as it is capable of detecting clusters of arbitrary shapes. This algorithm makes use of the spectrum (eigen values) of the affinity matrix (similarity matrix) of the data to perform dimensionality reduction before clustering in fewer dimensions. There are several ways to form our affinity matrix. Here, we used the simple method of finding the \emph{nearest neighbours}. 

However, the Spectral Clustering algorithm requires us to specify the optimal number of clusters before running the algorithm, which can be a little tricky, as there is no particular correct answer to this problem. Furthermore, the optimal number of clusters depend on the specified minimum number of elements that can be in a single cluster. 

Therefore, to solve this issue, we use the algorithm found in the work of \cite{inproceedings}. Using this method we calculate the optimal number of clusters for various values of the minimum number of elements allowed in a single cluster. The results are tabulated below in Table 1. The column values represents the minimum elements in a cluster, while the rows are the optimal number of clusters in decreasing order of significance. We have only considered the top 3 values, as the further we go down, the significance drops down considerably. Now, it is up-to human judgement to select the best combination from this set of values by trial and error. We handpicked the minimum elements in a single cluster and the number of clusters as 4 and 222 respectively, and performed Spectral Clustering of the anime titles. The results were added to the knowledge base.

\begin{table}[h!]\label{tab:1}
\caption{Optimal number of clusters for various values of minimum elements in a cluster}       
\begin{tabular}{c c c c c c c c c c c c c c c}
\hline\noalign{\smallskip}
1 & 2 & 3 & 4 & 5 & 6 & 7 & 8 & 9 & 10 & 11 & 12 & 13 & 14 & 15\\ [0.5ex]
\noalign{\smallskip}\hline\noalign{\smallskip}
45 & 100 & 71 & 91 & 83 & 66 & 68 & 56 & 59 & 58 & 65 & 64 & 40 & 33 & 39 \\
50 & 60 & 108 & 222 & 57 & 29 & 30 & 29 & 27 & 27 & 28 & 28 & 127 & 123 & 121\\
55 & 67 & 843 & 76 & 93 & 94 & 107 & 110 & 122 & 124 & 121 & 124 & 133 & 130 & 129 \\
\noalign{\smallskip}\hline
\end{tabular}
\end{table}

\subsubsection{Logically Opposite Cluster}  \label{Logically Opposite Cluster}
In certain cases, we need the logically opposite cluster of an anime title to make better recommendations. However, this can be a tricky problem, as there might not be an exact logically opposite anime to a particular anime. In such a case, we need to approximate our calculations. The process we used for finding the logically opposite cluster is summarised in the algorithm below.


\begin{algorithm}[H]
 
 1. Find the Centroid of all the anime titles $A$. Let it be $M$.\\
 2. Calculate 2 metrics, namely - \emph{d-metric} and \emph{cosine-metric}.\\
 3. To calculate the \emph{d-metric}, we first calculate the Euclidean distance of $a_0$ from $M$. Let it be $d_0$.\\
 4. For each anime title other the $a_0$, calculate their distances from $M$. Let them be $D = [d_1, d_2, d_3… d_n]$.\\
 5. For each distance measure $d_i$ from D, calculate the \emph{d-metric} = $d_i/(d_0 + e)$. The $e = 1e-9$ is for removing \emph{division by 0} error.\\
 6. If the \emph{d-metric} $ > 1$, we inverse it. Finally, we subtract it from 1. Now, the \emph{d-metric} lies between 0 and 1.\\
 7. For each anime title $a_i$ other than $a_0$, we calculate the cosine of the angle between the vectors joining $M$ with $a_i$ and $a_0$ using dot product of vectors. Then we add 1 to it. This gives us our \emph{cosine-metric}, which lies between 0 and 2. \\
 8. Finally, we add the corresponding \emph{cosine-metric} and the \emph{d-metric}. The minimum of which gives the logically opposite anime title of $a_0$. The cluster of this anime title is the logically opposite cluster of $a_0$.\\
 \caption{Logically Opposite Cluster}
\end{algorithm}

The opposite clusters for every anime title in the database is calculated and added to the knowledge base.

\subsubsection{Hybrid Filter}
\begin{figure}
    \centering
    \includegraphics[width=\textwidth]{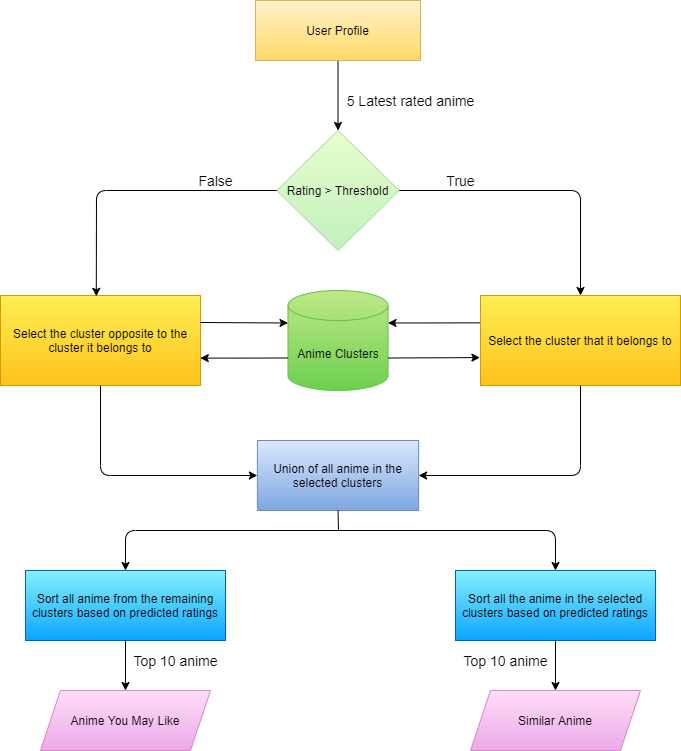}
    \caption{Novel Hybrid Filter}
    \label{fig:hybridfilter}
\end{figure}

The Hybrid Filter is named as such, because it combines the results of 2 different processes, namely - collaborative filtering and content-based filtering. The former is realised in our system by the \emph{Primary Autoencoder}, while the later is realised by the \emph{clustering} process. The way the filter works is described below. 
Let us denote the set of predicted ratings generated by the \emph{Primary Autoencoder} of all the anime titles not rated by the user yet as $S$. Furthermore, let $A = [a_1, a_2, a_3, ..., a_n]$ be the set of anime titles rated by the user in temporal order. It means that the anime title $a_1$ is rated in the beginning and the anime title $a_n$ is rated last. We consider the last 3 rated anime titles from $A$ to figure out the likes and dislikes of the user. Let the sets $C_1, C_2, C_3$ contain the anime titles in the clusters of the last 3 rated anime titles. Here, the clusters can be their own cluster or logically opposite one, based on the provided rating of the anime titles. We use a threshold value to make this distinction. Let their \emph{Union} be $C$. We take the intersection of the sets $S$ and $C$ and sort the anime titles in the resultant set in the decreasing order of their predicted ratings. The \emph{top 10} of them are tabulated and shown in the category \emph{Similar Anime}. For the \emph{Anime You May Like} category, we take the \emph{top 10} anime titles in the resultant set of $S - C$ after sorting them in decreasing order. These 2 lists are shown as recommendations to the user. This process is shown in the Fig. \ref{fig:hybridfilter}.

\subsection{User-System Interaction} \label{System Interaction}
RikoNet has been designed to follow a user throughout his entire journey of watching anime. The more the user watches anime and interacts with the system (adds his ratings), the better the system gets at recommendation. The system builds up the user profile as he rates more anime from the available dataset. While this setup ensures that the recommendations get better with time, RikoNet also manages to incorporate the changes in users' tastes and recommend accordingly.

When the user fires up the system for the first time, he is asked to enter his \emph{age} and \emph{sex}. This information is added to the \emph{user profile}. Then the user is asked if he would like to search and rate the anime titles which he has already watched prior to interacting with the system. The user can opt to skip this step. In this case, the user profile is empty and the system recommends purely based on anime titles popular in their respective genres. Otherwise, the system forms the recommendation lists of anime titles based on the current user profile. The user can further rate more anime titles from the recommendation lists themselves, or search and rate from the available dataset of anime titles. These new ratings are incorporated into the \emph{user profile} and are taken into account while forming the new recommendation lists. This process continues as long as the user demands, or he has not watched and rated every anime title present in the dataset.

\section{Experimental Setup and Results} \label{Results}

\subsection{Data Collection and Processing} \label{Data}
The data used for \emph{training} Rikonet was crawled from the popular anime website \href{https://myanimelist.net/}{MyAnimeList.net} using the \href{https://jikan.docs.apiary.io/#}{Jikan API}. The collected data was cleaned of duplicates as well as missing values and false data. Then it was fed into the algorithms that RikoNet uses.

\subsubsection{Dataset Contents}
The collected data consists of 3 types of entities which are described below:

\emph{1.} \textbf{\emph{Anime Titles:}} This dataset contains a list of anime, with both English and Japanese names, title synonyms, genre, studio, producer, duration, rating, score, airing date, episodes, source (manga, light novel, etc.), and other important data about individual anime, providing sufficient information on all the anime released till date. 

\emph{2.} \textbf{\emph{Users:}} This dataset contains information on users who watch anime, namely username, registration date (join date), last online date, birth date, gender, location, and lots of aggregated values from their anime lists(watch lists, also called \emph{User Lists}).

\emph{3.} \textbf{\emph{User Lists:}} This dataset contains data related to different users' personal experiences of various anime watched by them. Each record contains username, anime ID, score (rating out of 10), status(watched, dropped, plan-to-watch, etc.), and timestamp of when this record was last updated.

\subsubsection{Data Cleansing and Pre-processing}
Data cleansing is the first step of data pre-processing. First, we discarded those anime titles with an unknown studio or unknown source. Their ratings were removed too. Most of the removed anime were unknown and insignificant. After this process, we were left with 6668 unique anime titles with 43 genres in the whole dataset. Then we reduced the number of users by taking a threshold value for the number of anime watched and rated, removing all users who have rated less than that. After reduction, we were left with only 16,384 users in the whole dataset. Moreover, some of the users rated anime which they added to their \emph{"plan-to-watch"} lists. These data elements were considered as false data and discarded. Additionally, some users rated anime highly which were in their \emph{"dropped"} lists. These data elements were considered as \emph{erroneous} data and modified with the lowest possible rating, which is 1.

\begin{table}[h!] 
\caption{Anime-Genre sparse matrix}
\label{tab:genretable}       
\setlength{\tabcolsep}{3pt}
\resizebox{\textwidth}{!}{%
\begin{tabular}{c c c c c c c c c c c c}
\hline\noalign{\smallskip}
anime\_id & Sports & Parody & Mecha & Cars & Dementia & Slice of Life & Mystery & Horror & Super Power & Magic & ...\\
\noalign{\smallskip}\hline\noalign{\smallskip}
11013 & 0 & 0 & 0 & 0 & 0 & 0 & 0 & 0 & 0 & 0 & ...\\
2104 & 0 & 1 & 0 & 0 & 0 & 0 & 0 & 0 & 0 & 0 & ...\\
5262 & 0 & 0 & 0 & 0 & 0 & 0 & 0 & 0 & 0 & 1 & ...\\
721 & 0 & 0 & 0 & 0 & 0 & 0 & 0 & 0 & 0 & 1 & ...\\
\noalign{\smallskip}\hline
\end{tabular}}
\end{table}

Finally, the cleaned dataset was pre-processed to generate two sparse matrices - \emph{anime-genre sparse matrix} and \emph{user-anime rating sparse matrix}. Table \ref{tab:genretable} shows a glimpse of the former which was created from the \textbf{\emph{Anime Titles}} dataset.\par
For creating the \emph{user-anime rating sparse matrix} we used the \textbf{\emph{Users}} and \textbf{\emph{User Lists}} datasets. Here, Male is encoded as 0 and Female as 1. Furthermore, we categorized the users based on their age, which was calculated from the \emph{birth date} data entry . The categories are as follows -
\begin{align*}
    &Category 1: Age < 11\\
    &Category 2: 11 <= Age < 16\\
    &Category 3: 16 <= Age < 20\\
    &Category 4: 20 <= Age < 27\\
    &Category 5: Age >= 27
\end{align*}

\subsubsection{Training and Testing Dataset}
The final \emph{user-anime rating sparse matrix} with all the users categorized according to gender and age is used as the \emph{test dataset}. Each record holds the information of a user. All anime are represented by their respective \emph{anime id} forming individual columns. A rating of 0 to any anime title indicates that the user has not given any rating to it. Table \ref{tab:testset} shows a view of the test set.

The \emph{training dataset} is extracted from the test set. Here, 5\% of the non-zero ratings of each user are randomly converted to 0. When the model makes predictions for unrated anime, it will also predict ratings for these false zero ratings. Comparing these generated ratings with the original ones helps in evaluating how well the model is predicting.

\begin{table}[h!] 
\caption{Testing set}
\label{tab:testset}       
\setlength{\tabcolsep}{2pt}
\resizebox{\textwidth}{!}{%
\begin{tabular}{c c c c c c c c c c c c c c c c c c c}
\hline\noalign{\smallskip}
user\_id & Gender & Category2 & Category3 & Category4 & Category5 & 11013 & 2104 & 5262 & 721 & 12365 & 6586 & 178 & 2787 & 4477 & 853 & 4814 & 7054 & ...\\ 
\noalign{\smallskip}\hline\noalign{\smallskip}
4388263 & 0 & 0 & 0 & 1 & 0 & 0 & 0 & 0 & 0 & 0 & 0 & 0 & 0 & 0 & 0 & 0 & 0 & ...\\
6124996 & 0 & 0 & 0 & 1 & 0 & 0 & 0 & 0 & 0 & 0 & 0 & 0 & 0 & 0 & 0 & 0 & 9 & ...\\
463378 & 0 & 0 & 0 & 0 & 1 & 0 & 0 & 0 & 0 & 0 & 0 & 0 & 0 & 0 & 0 & 0 & 10 & ...\\
88520 & 0 & 0 & 0 & 0 & 1 & 0 & 0 & 0 & 0 & 0 & 0 & 0 & 0 & 0 & 0 & 0 & 0 & ...\\
142000 & 0 & 0 & 0 & 0 & 1 & 0 & 0 & 0 & 0 & 0 & 0 & 0 & 5 & 0 & 0 & 0 & 8 & ...\\
\noalign{\smallskip}\hline
\end{tabular}}
\end{table}

\subsection{Performance Parameters} \label{Parameters}

As an optimizing criteria for the model, two evaluation metrics are used, namely, \emph{Mean Squared Error (MSE)} and \emph{Root Mean Squared Error (RMSE)}.

\begin{equation}
     MSE = \frac{\sum_{i=1}^{n}m_{i}\times (r_{i}-y_{i})^2}{\sum_{i=1}^{n}m_{i}}
\end{equation}
\begin{equation}
    RMSE = \sqrt{\frac{\sum_{i=1}^{n}m_{i}\times (r_{i}-y_{i})^2}{\sum_{i=1}^{n}m_{i}}}
\end{equation}
\begin{align*}
    Here,\ &r_{i}\ is\ actual\ rating \\
    &y_{i}\ is\ predicted\ rating \\
    &m_{i} = mask\ applied=\left\{\begin{matrix}
             0,& r_i = 0\\ 
             1,& otherwise 
            \end{matrix}\right.
\end{align*}

\subsection{Effects of various activation functions}

We trained our model using different activation functions to determine the one which gives best results in this scenario. The most commonly used functions in deep learning were explored : \emph{rectified linear units (ReLU), exponential linear units (ELU)} - \cite{clevert2015fast}, \emph{hyperbolic tangent (tanh)}, and \emph{scaled exponential linear units (SELU)} - \cite{klambauer2017self}. Since all ratings are within the range from 1 to 10, ReLU activation is applied in the last layer to trim out the negative predicted ratings.

\begin{figure}
    \begin{subfigure}{.8\textwidth}
    \centering
    \includegraphics[width=1\linewidth]{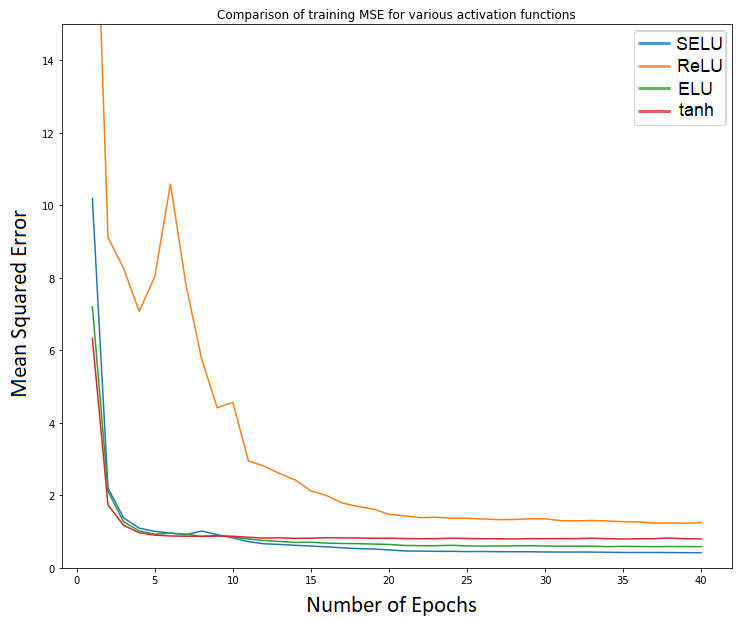}
    \caption{Training}
    \end{subfigure}
    \begin{subfigure}{.8\textwidth}
    \centering
    \includegraphics[width=1\linewidth]{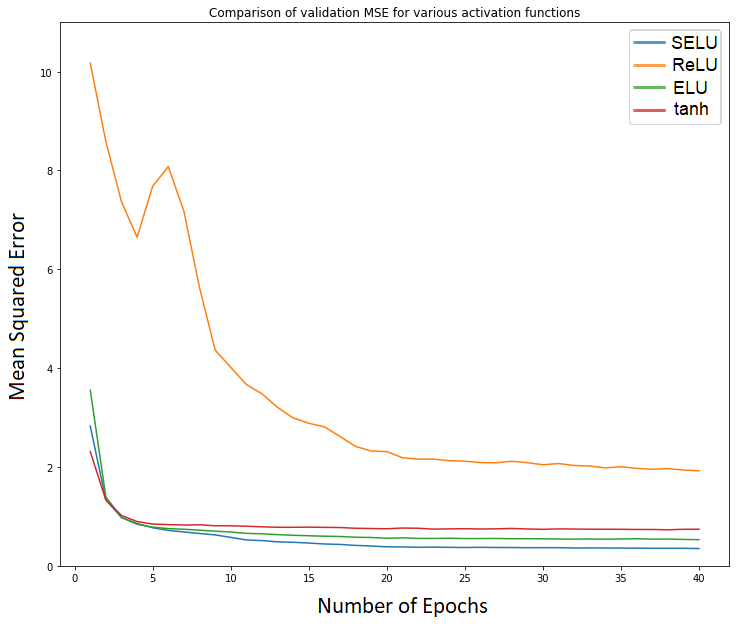}
    \caption{Validation}
    \end{subfigure}
    \caption{Comparing MSE for ReLU, tanh, ELU and SELU activation functions}
    \label{fig:activationComparison}
\end{figure}

Comparing the MSE for these functions revealed that SELU, ELU and tanh activations performed much better than ReLU activation. Moreover, the former two performed significantly better than the other two. The experiment inferred that the two properties: a) unbounded positive part, and b) non-zero negative part, play a significant role in this case of training the autoencoder. Hence SELU was chosen as activation function which performed better than all other functions.
Figure \ref{fig:activationComparison} shows the performance comparisons of these functions.

\section{Discussions and Analysis} \label{Discussions}

\subsection{Comparative Evaluation} 
We tested our model on the MovieLens 100K dataset and performance is shown in the Table \ref{tab:errortable}. Performance of different other models developed till now, on the same dataset is also enlisted in the table for comparison. All the values for other techniques are cited from the paper by \cite{kiran2020dnnrec}, who collected these from various other papers for comparison. Both MSE and RMSE of RikoNet are \emph{18.267\%} and \emph{9.63\%} better than the previous best technique respectively. As we can see, RikoNet outperforms the previous techniques by a large margin. 

\begin{table}[h!]
    \centering
    \caption{Comparative analysis  of MSE and RMSE measures}
    \begin{tabular}{|c|c|c|}
    \hline
    \textbf{Method} & \textbf{MSE (lower is better)} & \textbf{RMSE (lower is better)} \\
    \hline
    Global Average & 1.129 & 1.062  \\
    \hline
    User Average & 0.929 & 0.964 \\
    \hline
    User KNN – Cosine & 0.813 & 0.813 \\
    \hline
    Item KNN - Cosine & 0.854 & 0.862 \\
    \hline
    User KNN – Pearson & 0.814 & 0.902 \\
    \hline
    Item KNN - Pearson & 0.796 & 0.892 \\
    \hline
    SVD & 0.879 & 0.937 \\
    \hline
    Biased MF & 0.768 & 0.877 \\
    \hline
    SVD++ & 0.756 & 0.869 \\
    \hline
    DNNRec \cite{kiran2020dnnrec}  & 0.747 & 0.864 \\
    \hline
    UBP \cite{9357332}  & 0.724 & 0.851 \\
    \hline
    PDMF-post 50D \cite{yuan2021preliminary}  & 0.427 & 0.654 \\
    \hline
    \textbf{Proposed Model: RikoNet} & \textbf{0.349} & \textbf{0.591} \\
    \hline
    \end{tabular}
    \label{tab:errortable}
\end{table}

\subsection{Human Evaluation}

\subsubsection{Test Design and Overview}
We reached out to 20 domain experts to evaluate our system. Each of them had varied tastes and were from different walks of life. We contacted them through personal mail using our connections.

Among the 20 study participants, half of them were males and the other half were females. Their age varied from 18 to 30 years old. In each recommendation, we showed them 10 \emph{Similar Anime} and 10 \emph{Anime You May Like}. 

At the beginning of each session, we asked the users to provide their age and gender to create their profiles. Following that, they were asked to rate at least 5 anime titles of their choice to get the engine started. The accepted ratings were from 1 to 10 integer values. Then the system showed them a recommendation list, which the users could watch and explore using other third party platforms. Having done that they could either rate or ignore those anime titles. Finally, the users were asked to rate the entire recommendation list out of 10, with 10 being the highest and 1 being the lowest.

\subsubsection{Recommendation Quality over Time}

\begin{tikzpicture}
\begin{axis}[
    title={Rating of the recommended list over time},
    xlabel={Iterations},
    ylabel={Average User Rating},
    xmin=0, xmax=10.5,
    ymin=0, ymax=10,
    xtick={0,1,2,3,4,5,6,7,8,9,10},
    ytick={0,1,2,3,4,5,6,7,8,9,10},
    ymajorgrids=true,
    grid style=dashed,
]

\addplot[
    color=blue,
    mark=diamond,
    ]
    coordinates {
    (1,7.2)(2,7.61)(3,7.57)(4,7.92)(5,8.13)(6,8.41)(7,8.37)(8,8.77)(9,8.68)(10,8.97)
    };
\end{axis}
\end{tikzpicture}

Over time, the amount of ratings provided by a particular user increases as he uses the system. These ratings are added to the user's profile, which are used to further improve the predictions. Hence, the quality of the recommendations are expected to improve as well. The system is also able to adapt to changes in the users' tastes, thereby personalizing the recommendations to the particular user. This should be reflected in the users' score of the recommendation lists. The above graph shows the average of the accumulated ratings from the participants of the test. Majority of the users ran the system for 10 iterations (successive recommendations). The graph clearly reveals the steady increase in score from $7.2$ to $8.97$ as more lists are worked through by the users, hence proving the credibility of our system.

\section{Conclusions and Future Scope} \label{Conclusion}
Anime is a diverse medium of entertainment with varying genres and themes targeting all demographic. This makes the task of making personal anime recommendations a formidable endeavour. After careful considerations, we have developed a novel anime recommendation engine to address this issue. RikoNet uses a new and tuned hybrid filtering approach combining the results of autoencoders and spectral clustering to adapt to the users' preferences. Usage of deep neural network has boosted the rating prediction capability of RikoNet outperforming the current state-of-the-art techniques. The system is also capable of recommending serendipitous anime titles for the user to explore.

Our experiments with RikoNet, however, reveal some shortcomings in the system. Since we are taking the three latest rated anime titles for finding the relevant clusters, a possibility arises wherein all three anime titles fall in a single cluster. This particular cluster may not contain a significant number of anime titles to further recommend. Moreover, this case of using the latest rated anime for recommendation can make the system very skewed in it's judgements. Apart from this, RikoNet requires the user to actively rate from the \emph{Anime You May Like} list for it to show a good variation. All of these bring up an opportunity for future research that can improve the system further.

\section {Declarations}

\textbf{Conflict of interest:} The authors declare that they have no conflict of interest.\\\\
\textbf {Availability of data and material:} The prototype of RikoNet is available online with an introductory tutorial on how to obtain anime recommendations, in our github repository  \url{https://github.com/NilutpalNath/RikoNet}.

\bibliographystyle{spbasic_updated}
\bibliography{main}

\end{document}